\documentclass[12pt]{article}
\usepackage{amssymb}
\usepackage{amsfonts}
\usepackage{amsmath}
\usepackage{graphicx}

\def\DJ{\leavevmode\setbox0=\hbox{D}\kern0pt
 \rlap{\kern.04em\raise.188\ht0\hbox{-}}D}
\def\dj{\leavevmode\setbox0=\hbox{d}\kern0pt
 \rlap{\kern.215em\raise.46\ht0\hbox{-}}d}

\begin{document}

\title{Consensus formation on simplicial complex of opinions}
\date{}
\author{Slobodan Maleti\' c\thanks{supersloba@vinca.rs}\\
Milan Rajkovi\' c\thanks{milanr@vinca.rs}\\
Institute of Nuclear Sciences Vin\v ca, University of Belgrade,\\
 Belgrade, Serbia}
\maketitle

\begin{abstract}
Geometric realization of opinion is considered as a simplex
and the opinion space of a group of individuals is a simplicial
complex whose topological features are monitored in the process
of opinion formation. The agents are physically located on the
nodes of the scale-free network.
Social interactions include all concepts of social dynamics present
in the mainstream models augmented by four additional interaction
mechanisms which depend on the local properties of opinions and
their overlapping properties. The results pertaining to the formation
of consensus are of particular interest. An analogy with quantum
mechanical pure states is established through the application of the high
dimensional combinatorial Laplacian.

\end{abstract}

\section{Introduction}

The interest of physics community in modeling problems of social dynamics,
in particular the application of statistical physics concepts and methods on
the modelling of opinion formation and dynamics gave birth to many simple
opinion models \cite{Castellano}. We mention here just those which had major
influence on advances in this field such as the Voter model \cite{Clifford},
Galam model \cite{Galam}, social impact model \cite{Nowak}, Sznajd model
\cite{Sznajd}, Deffuant model \cite{Deffuant}, and the Krause-Hegselman
model \cite{Hegselman}. Some of these models consider population of
individuals (agents) with discrete opinions represented as integers, for
example (+1/-1) \cite{Clifford}, \cite{Galam}, \cite{Nowak}, \cite{Sznajd},
while others consider population of individuals with continuous, bounded
range of opinions \cite{Deffuant}, \cite{Hegselman}. In the so called
consensus models \cite{Castellano} computer simulation of opinion dynamics
starts with Monte Carlo simulation of randomly distributed opinions over the
population of agents located on the nodes of a graph. The graph types
include small world networks \cite{Watts}, scale-free networks \cite%
{Barabasi}, and fully connected graphs enabling each agent to interact with
every other agent \cite{Deffuant}. Interaction between agents differs from
model to model and at the end of simulation leads to the state which may be
characterized as consensus (single opinion state), polarization (two
opinions), or anarchy (diversity of opinions). The majority of models are
characterized by interactions between agents which may include social
phenomena characterizing realistic communication situations such as
transmission of information from the individual to his(her) neighbors,
social influence, homophily and bounded confidence \cite{Castellano}.

The social actions of an individual reflect his (her) opinions as systems of
beliefs on different issues (or generally subjects in their broadest
meaning) which may or may not have an empirical background. In this sense
the opinions involve an individual's perception and knowledge or emotional
conditions about the likelihood of events or relationships regarding some
topic, and they also may involve evaluations of an event or an object \cite%
{Oskamp}. An individual may have an opinion about different issues, such as
religious matters, treatment of criminals or expansion of crime, political
issues including political parties, advertising, movies, ... or nearly
anything. In order to formulate or express an opinion on certain topic an
individual evaluates a set of interconnected judgements. When, for example,
a person talks, writes a mail or an article about his(her) opinion, he(she)
actually sends a "package" of interconnected judgements which form an
opinion. In this sense an opinion is not a collection of separate but
interrelated and interconnected judgements which form an authentic opinion.
Let us illustrate this with an example considering the issue of crime
expansion. An individual forms an opinion on the crime expansion by forming
and expressing judgements on topics of \textit{organized crime, corruption}
and \textit{gambling}, to name only a few causes responsible for the
increase of crime rate. When the conversation topic is the crime rate a
person expresses his opinion using all three judgments in an interconnected
manner and juxtaposes them under a certain relationship. Adding a new
judgement on a possible crime cause, like \textit{low incomes}, a person's
opinion shifts to the new opinion since now four judgements are
interconnected and interrelated. Of course, all four mentioned judgements
can be treated as opinion issues on their own, however we will not discuss
this case here since they are used here only for illustration purposes.

Collection of opinions of a large number of individuals represents an
entanglement of overlapping opinions and shared judgements, the analysis of
which requires a suitable mathematical framework which captures the essence
of opinions and their formation. Let us call the set of all different
opinions the opinion set and the set of all judgements the judgement set.
One approach would consider these as two defining sets of bipartite graph.
In this case the judgements are represented as vertices, and two vertices
are connected by a bond if two corresponding judgements are part of the same
opinion. Analyzing this structure from the graph-theoretic point of view the
actual information about opinions is lost since only connectivities between
judgements are considered. We can keep track of judgments as the basic
constituent elements of opinions and their mutual relationships however the
information about each opinion as a whole is unavailable. An alternative and
more comprehensive way is to consider opinion and judgement sets from the
point of view of combinatorial algebraic topology. In this approach the
judgements are again represented as vertices, opinions are now represented
as simplices \cite{Maletic}, while the individuals remain located on the
nodes of the scale-free network. A \textit{n}-simplex represents an \textit{n%
}-dimensional polytope which actually is the convex hull of its n + 1
vertices. Hence, a 2-simplex is a triangle, a 3-simplex is a tetrahedron and
so on. A simplex is not only a graph but also includes higher dimensional
faces and a space enclosed by the faces. Since simplices overlap by sharing
vertices the collection of simplices together with their overlaps is called
simplicial complex. The graph itself is a $1$-dimensional simplicial
complex. The simplicial complex has a three-fold mathematical property since
it can be considered from a topological, combinatorial and algebraic aspect
and may be most efficiently analyzed using concepts combining these three
mathematical disciplines. We consider the geometric realization of an
opinion as a simplex and the opinion set as a simplicial complex formed
through the overlapping connectivity structure of opinions which is realized
through shared judgements. The simplicial complex of opinions also encloses
an opinion space defined by the complex itself. The number of judgements
that characterizes a single opinion need not be the same and it may vary
from person to person. An example of two persons having opinions on the
expansion of crime, represented geometrically as simplices, is presented in
Figure 1. The case when they share certain judgements is presented on the
left and the case when they do not share any judgements is presented on the
right.

\begin{center}
\includegraphics[width=70mm]{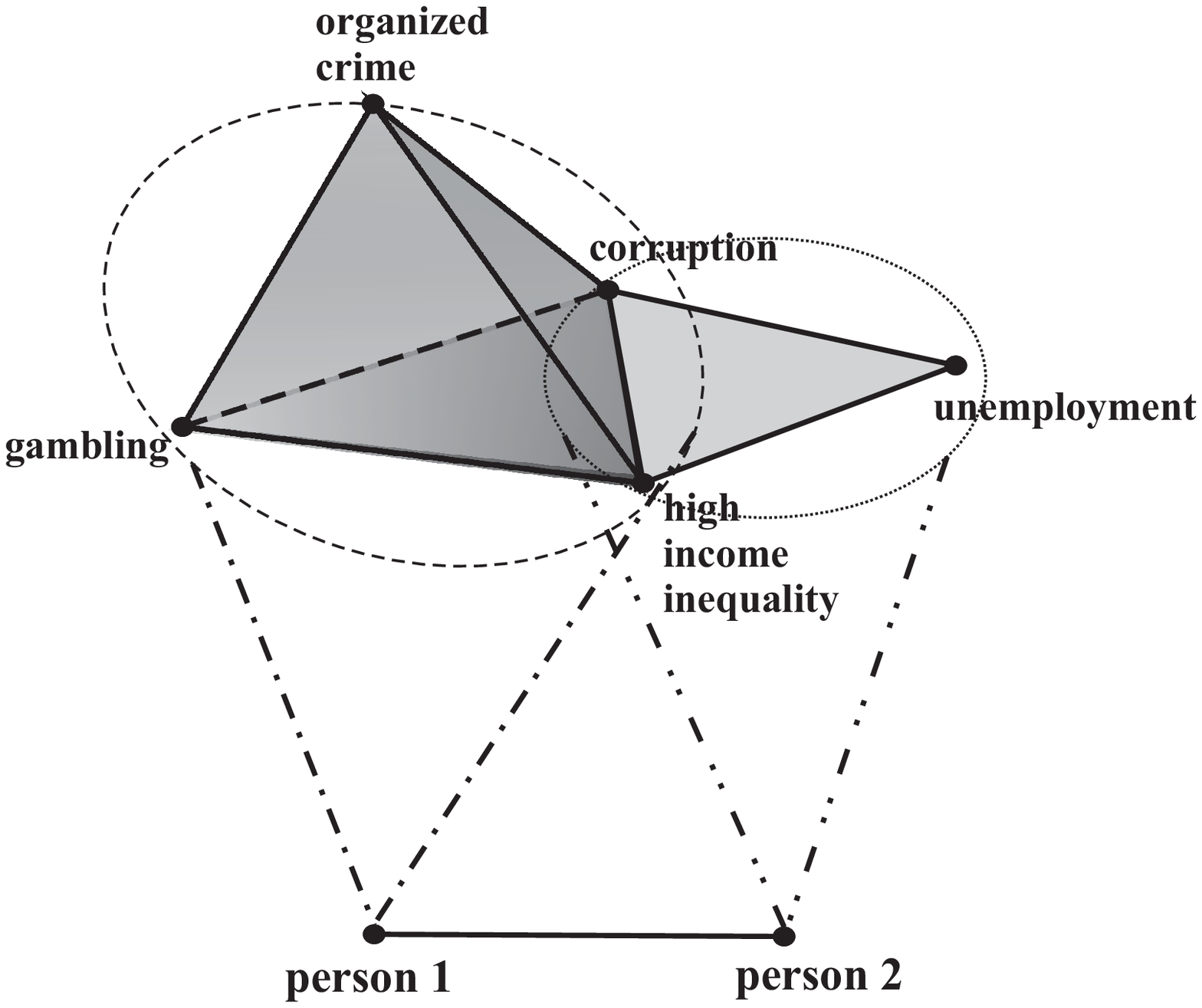}\hskip1mm %
\includegraphics[width=75mm]{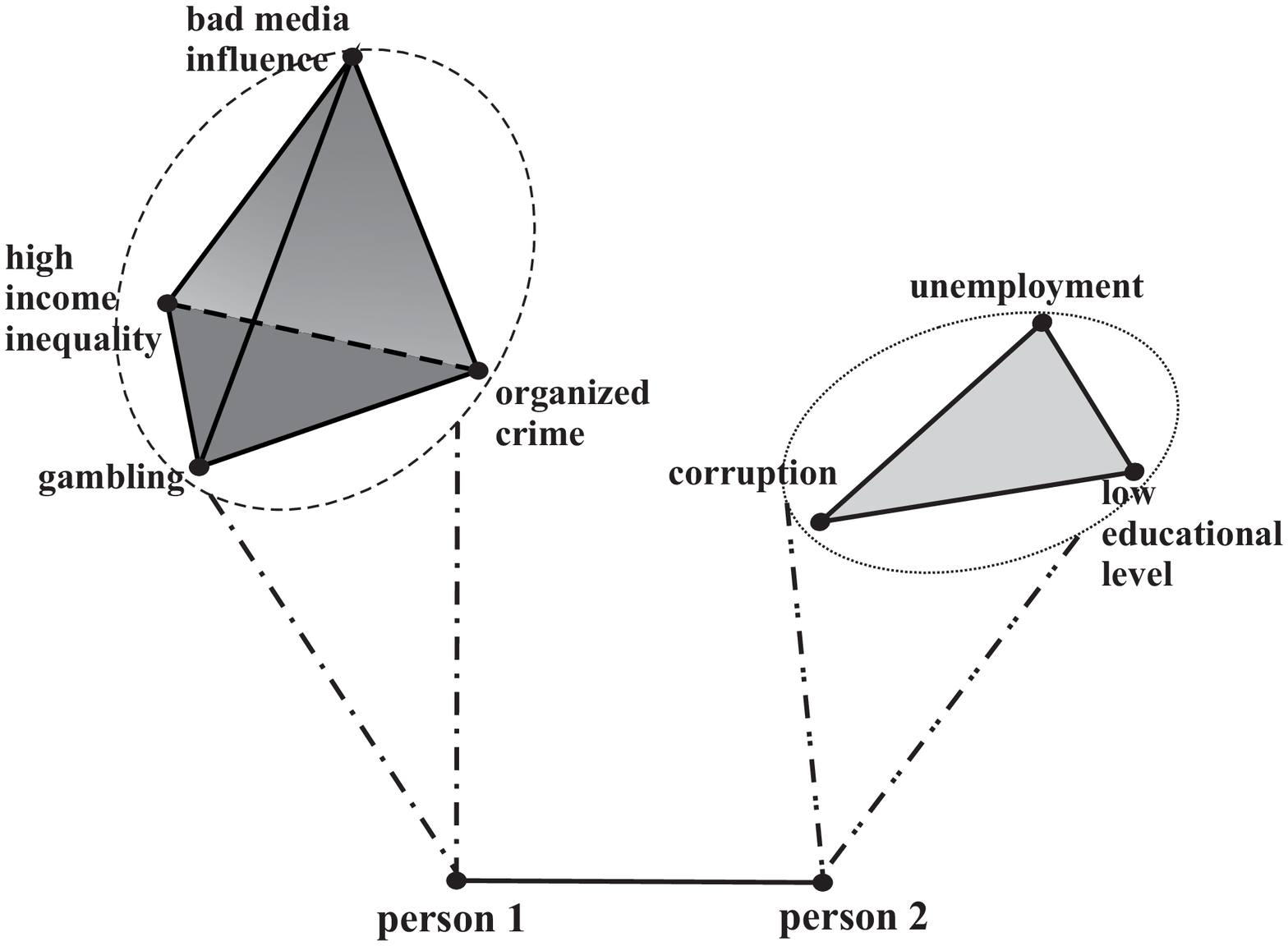}

Figure 1 Two persons expressing opinion on the expansion of crime. The case
of overlapping judgements is on the left and the case of non overlapping on the right.
\end{center}

Communication between individuals alters the shape of the opinion space and
by numerically evaluating certain topological quantities and topological
invariants \cite{Maletic} these changes can be monitored and predictions
about the final state can be made, for example, whether the consensus will
be achieved or not. The predictions are further supported by the existence
of pure opinion states defined through the properties of properly normalized
Combinatorial Laplacian as explained further in the text.

Social entities in the form of individuals (or agents) are the carriers of
social interpersonal interaction \cite{Krech} so that the opinion of a
single person or a group of persons may be under the influence of another
individual or another group of individuals. Interactions may be of dyadic
type (between two persons), triadic type (between three persons) or in
general of\ \textit{n}-adic type which includes \textit{n} persons and
clearly social interaction depends on the interpersonal connectivity of
persons. In the model presented here social interactions include all
concepts of social dynamics present in the mainstream models such as
information flows directed outwards (i.e. from the agent to its neighbors),
social influence, homophily and bounded confidence. There are four
additional interaction and communication activities which may alter
individual's opinion in this model. The result of the interaction to a large
extent depends on whether two agents have the same or diverging opinions, or
if they have different but to a certain extent overlapping opinions. If two
agents have the same opinion they convince all their common neighbors to
adopt this opinion. This type of interaction is in common with the
interaction of the Sznajd model \cite{Sznajd}. If two agents have different
but overlapping opinions, the result of their interaction depends on the
extent of the overlap, and can be twofold. The agents can either exchange
one of the judgements forming compromise or they can unite their opinions by
uniting judgements (forming joint opinion) and the probability of the two
outcomes is proportional to the extent of the overlap. In both cases new
opinions are formed and added to the opinion space. This type of interaction
is similar to the Axelrod model \cite{Axelrod} with some important
differences. In Axelrod model the culture is defined as a vector $F=({f_{1}},%
{f_{2}},...,{f_{F}})$ and each entry $f_{i}$ (called cultural feature) can
take a value from the so called trait set ${f_{i}}\in \{{\tau _{1}},{\tau
_{2}},...,{\tau _{T}}\}$, whereas in our model the opinion represented as
simplex, is an unordered set of judgements. Furthermore, the length of
vector $F$ in the Axelrod model is fixed while in our model the dimension of
the opinions (number of judgements which characterize them) may vary. With
respect to this property our model is somewhat closer to the naming game
model \cite{Baronchelli}, \cite{Dall'Asta}, although the tendency in this
model is to decrease the inventory of word-associations to common words,
while in our case the tendency is to increase the number of judgements which
characterize the opinion. The opinions in our model may experience
alteration in one additional way. The model allows for a new previously
absent judgement to be added to the judgements set under specific
conditions. Consequently, the agent's corresponding opinion changes and the
new opinion is added into the complete set of opinions. This type of
interaction mechanism has remote resemblance to the "cultural drift"
mechanism introduced as a variant of the Axelrod model in \cite{Klemm},
however the possibility to change the trait spontaneously is driven by
random noise and in our model the change is caused by the local property of
agent's opinion.

One of the aims of this study is also to promote the model founded on the
structure of simplicial complex as the natural setting for opinion formation
and other social network studies which offers a number of new aspects and
results in comparison with the standard graph structure approach. The
organization of the paper is as follows. Following a brief introduction to
topological and geometrical features of simplicial complexes in Section 2,
we present the model of opinion formation on simplicial complexes and the
algorithm in a detailed manner in Section 3. The results of the simulation
are presented in Section 4 and discussion of results and conclusion is
presented in Section 5.

\section{Simplicial complex}

\subsection{Definition and topological features}

A short introduction to the topology of simplicial complexes presented in
this Section covers basic properties and fundamental topological measures
used in the analysis. More detailed description of terms and concepts used
may be found in \cite{Maletic}, \cite{Mi 1} and \cite{Mi 2}.

Simplicial complexes are formed by simplices which may have different
dimensions, and hence can be analyzed as multidimensional and multilevel
objects. Given a set of points whose elements we call \textit{vertices} $V=\{%
{v_{0}},{v_{1}},...,{v_{n}}\}$, any subset of $q+1$ elements of this set $%
\{v_{\alpha _{0}},v_{\alpha _{1}},...,v_{\alpha _{q}}\}$ is called \textit{%
q-dimensional simplex}, or simply \textit{q-simplex}. A $p$-simplex $\sigma
_{p}$ is a \textit{p-face} of a $q$-simplex $\sigma _{q}$, denoted by $%
\sigma _{p}\leq \sigma _{q},$ if every vertex of $\sigma _{p}$ is also a
vertex of $\sigma _{q}$, and if two simplices $\sigma _{q}$ and $\sigma _{r}$
have $p+1$ common vertices then they share a $p$-face. A simplicial complex
represents a collection of simplices together with their faces. In more
formal terms a simplicial complex $K$ on a finite set $V=\{v_{1},...,v_{n}\}$
of vertices is a nonempty subset of the power set of $V$, so that $K$ is
closed under the formation of subsets \cite{Munkres}. The maximal dimension
of a simplex in $K$ determines the dimension of the whole simplicial
complex. A graph is a 1-dimensional simplicial complex with the same set of
vertices and with simplices represented as edges of a graph. An illustration
of the simplicial complex is presented in Figure 2.

\begin{center}
\includegraphics[width=70mm]{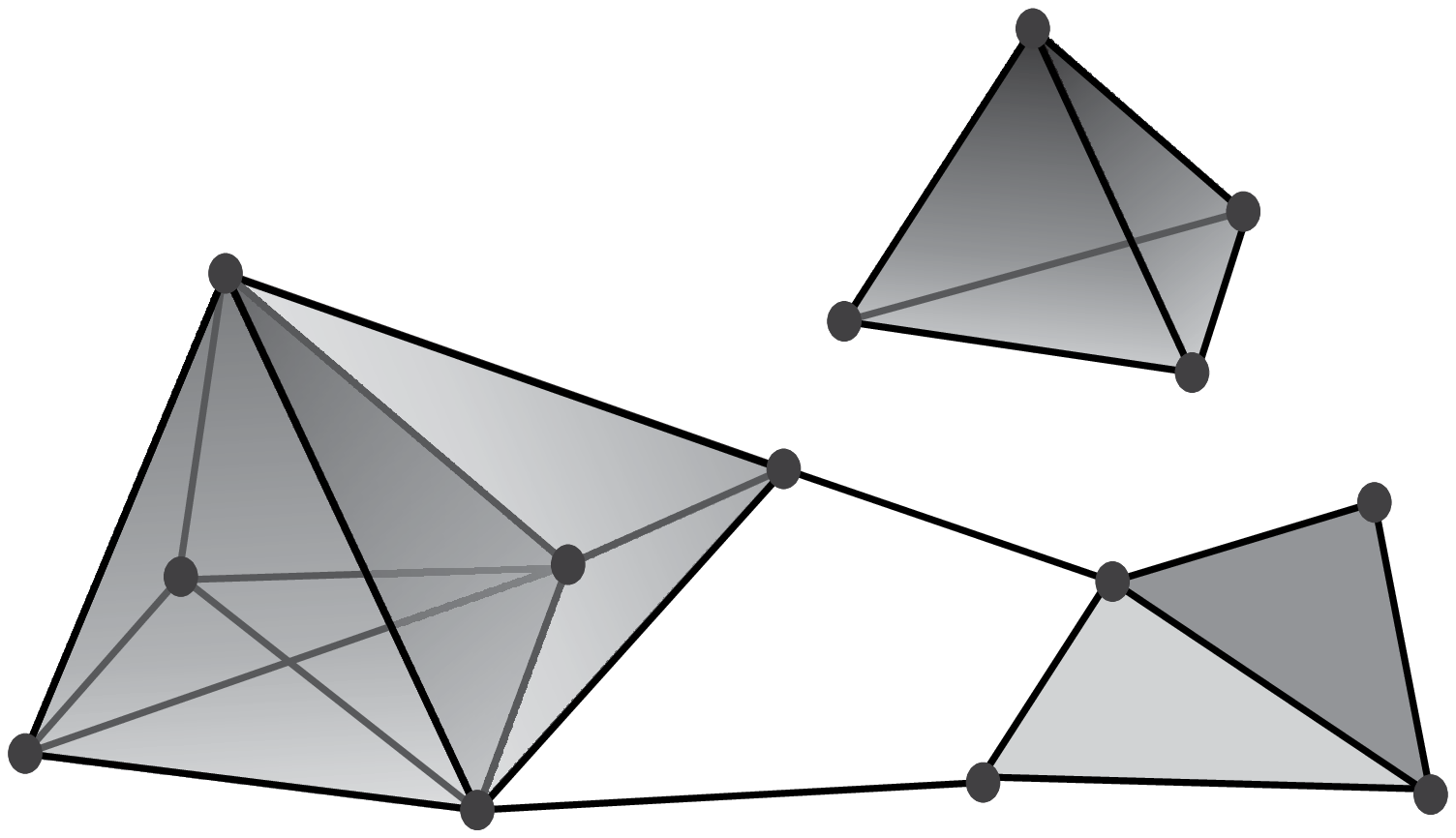}

Figure 2 An illustration of simplicial complex formed by one 4-simplex, two 3-simplices, two 2-simplices, and two 1-simplices.
\end{center}

Throughout this
presentation we adopt the notation where the subscript marks the dimension
of a simplex and the superscript marks the numerical name of a simplex,
hence the notation $\sigma _{q}^{i}$ means the "$q$-dimensional simplex $i$%
". Two simplices $\sigma $ and $\rho $ are $q-connected$ \cite{Johnson} if
there is a sequence of simplices $\sigma ,\sigma ^{1},\sigma ^{2},...,\sigma
^{n},\rho ,$ such that any two consecutive ones share at least a $q$-face.
The $q$-connectivity between simplices induces an equivalence relation on
simplices of a complex $K$, since it is reflexive, symmetric, and
transitive. This equivalence relation be denoted by $\gamma _{q}$ so that
\begin{equation*}
(\sigma ^{i},\sigma ^{j})\in \gamma _{q}\text{ \ \ \ \ \ if and only if }%
\sigma ^{i}\text{ is q-connected to }\sigma ^{j}.
\end{equation*}%
Let $K_{q}$ be the set of simplices in $K$ with dimension greater than or
equal to $q$. Then $\gamma _{q}$ partitions $K_{q}$ into equivalence classes
of $q$-connected simplices. These equivalence classes are members of the
quotient set ${K_{q}}/{\gamma _{q}}$ and they are called the \textit{$q$%
-connected components} of $K$. Every simplex in a $q$-component is $q$%
-connected to every other simplex in that component, but no simplex in one $%
q $-component is $q$-connected to any simplex on a distinct $q$-connected
component. Changing the roles of simplices and vertices of simplicial
complex $K$ new simplicial complex $K^{-1}$ is formed in which simplices are
the vertices of $K$ and vertices are the simplices of $K$. This new
simplicial complex $K^{-1}$ is called \textit{the conjugate complex} of
simplicial complex $K$ \cite{Johnson}. Illustration of the original and the
conjugate complex is presented in Figure 3 (a) and (b), respectively.
Vertices are labeled by numbers (Figure 3(a)) and letters (Figure 3(b)),
whereas simplices by $\sigma (i)$, where $i=a,b,c,...$ (Figure 3(a)) and $%
i=1,2,3,...$ (Figure 3(b)).

\begin{center}
\includegraphics[width=140mm]{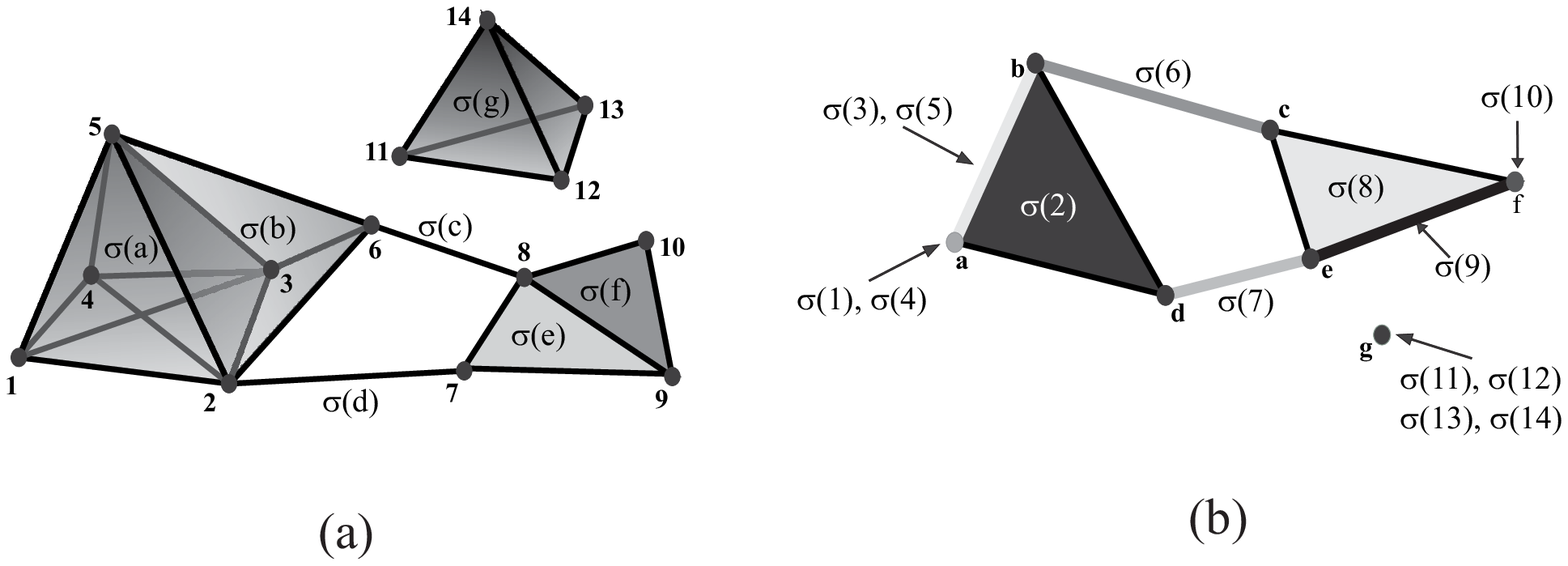}

Figure 3 An illustration of simplicial complex (a) and its conjugate (b).
\end{center}

\subsection{Structure properties of simplicial complexes}

\textbf{Q-vector (first structure vector).} The cardinality of ${K_{q}}/{%
\gamma _{q}}$ is denoted $Q_{q}$ and is the number of distinct $q$-connected
components in $K$. The value $Q_{q}$ is the $q^{th}$ entry of the so called
\textit{Q-vector} (\textit{first structure vector}) \cite{Johnson}, an integer vector with
the length $dim(K)+1$. An example illustrating the partitioning the
simplicial complex into $q$-connectivity classes and Q-vector is presented
in Figure 4. Hence,
Q-vector describes the structure of simplicial complex on different levels
of connectivity. The notation is \cite{Johnson}:
\begin{equation}
\mathbf{Q}=({Q_{(dim(K))}},{Q_{(dimK-1)}},...,{Q_{0}})\,.
\end{equation}

\begin{center}
\includegraphics[width=70mm]{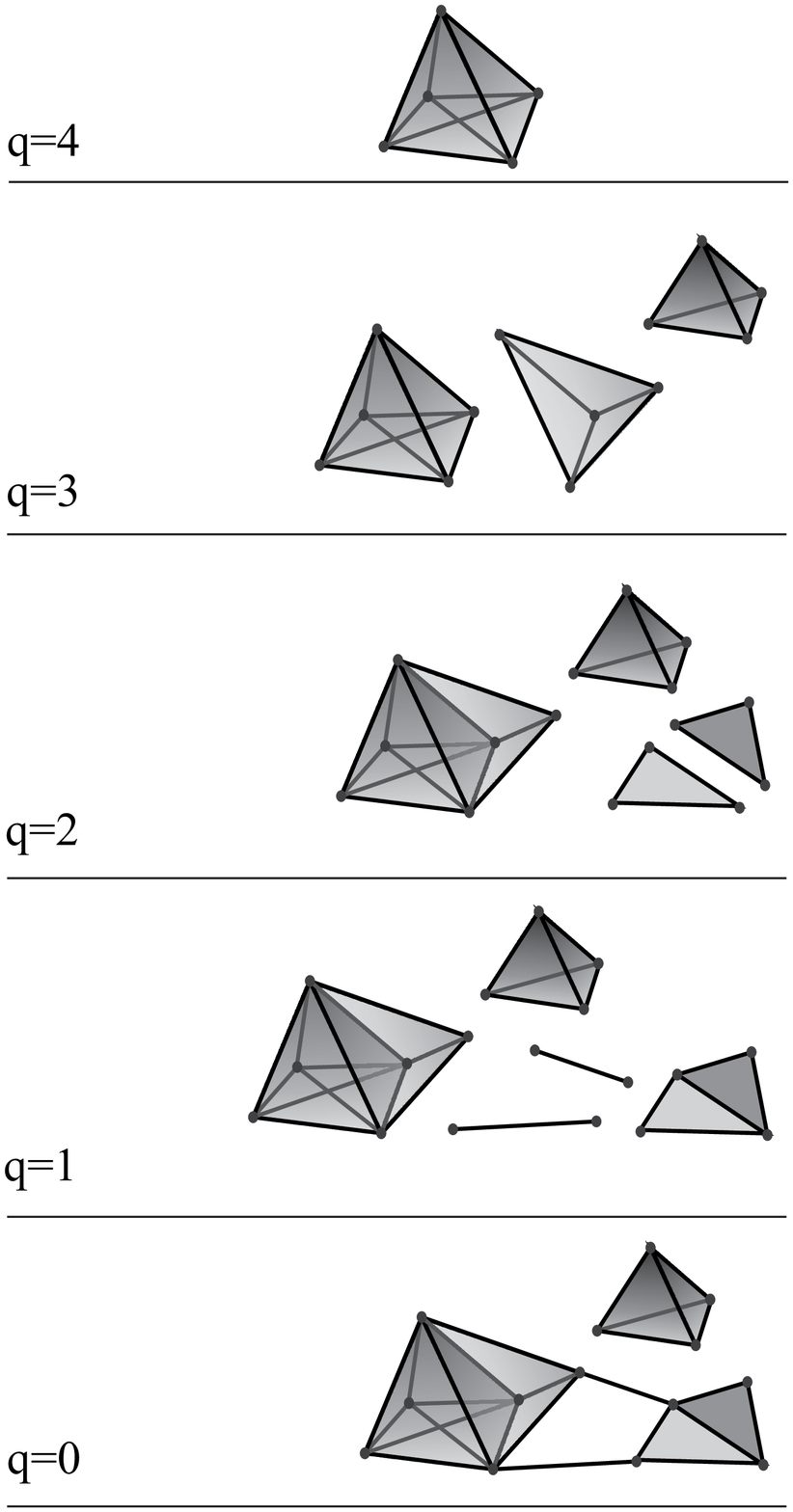}

Figure 4 Graphical illustration of the Q-vector of simplicial complex from Figure 2.
\end{center}

\textbf{Eccentricity.} Define $\hat{q}$ (top q) as the dimension of the
simplex and define $\check{q}$ (bottom q) as the largest dimension of faces
which simplex $\sigma$ share with other simplices, i.e. the largest $q$%
-nearness value. $\check{q}$ is equivalent to the value of the $q$-level on
which the simplex firstly connects to some other simplex. Then the
eccentricity of simplex $\sigma$ is defined as \cite{Maletic}:
\begin{equation}
ecc(\sigma)={\frac{{\hat{q}-\check{q}}}{{\hat{q}+1}}}\, .
\end{equation}

$ecc(\sigma)$ measures the individuality of a simplex, and indicates degree
of integrity of the simplex $\sigma$ in a simplicial complex. The simplex
which has $ecc=0$ is completely integrated into the structure, i.e. the
simplex is face of another simplex. The simplex which has $ecc=1$ does not
share vertices (faces) with any other simplex, i.e. it is completely
disintegrated.

\textbf{Vertex significance of a simplex.} One vertex can be part of many
simplices and a vertex weight $\theta $ provides information on the number
of simplices which are created by that vertex. The sum of weights of the
vertices which create simplex ${\sigma _{q}}(i)$ yields $\Delta ({\sigma _{q}%
}(i)),$ and the vertex significance of the simplex is defined as \cite%
{Maletic}:
\begin{equation}
vs({\sigma _{q}}(i))={\frac{{\Delta ({\sigma _{q}}(i))}}{{{max_{k}}\Delta ({%
\sigma _{q}}(k))}}}\,,
\end{equation}%
where ${max_{k}}\Delta ({\sigma _{q}}(k))$ is the maximal value of all ${%
\sigma _{q}}(i)$. The larger value of $vs$ indicate larger importance of the
simplex with respect to the vertices which create it in the sense that,
compared to other simplices, it contains vertices which take part in the
construction of the larger number of other simplices.

\textbf{Combinatorial Laplacian.} The matrix representation of the $q^{th}$
\textit{Laplacian matrix} of simplicial complex $K$ is
\begin{equation*}
\mathcal{L}_{q}=B_{q+1}B_{q+1}^{T}+B_{q}^{T}B_{q}\,.
\end{equation*}
The boundary operator which maps simplices of dimension $q+1$ to simplices
of dimension $q$ is represented by the matrix $B_{q}$ so that the rows of $%
B_{q}$\ are associated with simplices of dimension $q+1$\ and columns are
associated with simplices of dimension $q$. Graph represents a $1$%
-dimensional simplicial complex since links ($1$-dim simplices) connect
nodes ($0$-dimensional simplices) and the largest dimension of a simplex in
the complex is $1$. In $B_{1}$, the matrix representation of the boundary
operator $\partial _{1},$ the rows are associated with edges and columns are
associated with vertices so that the matrix $B_{1}$ is equal to the
incidence matrix of an oriented graph. Therefore matrix representation of
the combinatorial Laplacian is ${\mathcal{L}_{0}}={B_{1}}{B_{1}^{T}}$ and
the matrix elements are
\begin{equation}
{({\mathcal{L}_{0}})_{ij}}=%
\begin{cases}
deg({v_{i}}), & \text{if $i=j$} \\
-1, & \text{if ${v_{i}}\sim {v_{j}}$} \\
0\ , & \text{otherwise}%
\end{cases}%
\end{equation}%
where $deg({v_{i}})$ is vertex degree (that is number of neighbors of a
vertex $v_{i}$) and the relation ${v_{i}}\sim {v_{j}}$ is the adjacency
relation between vertices $v_{i}$ and $v_{j}$. Clearly, the entries of the
0-dimensional combinatorial Laplacian are the same as the graph Laplacian
entries defined in the usual way via expression ${L_{qraph}}=D-A$, where
diagonal entries of matrix $D$ are equal to the vertex degrees (${D_{ii}}%
=deg({v_{i}})$) and nondiagonal entries are zeros, and the entries of matrix
$A$ are ${{(A)}_{ij}}=1$ if ${v_{i}}\sim {v_{j}}$, ${{(A)}_{ij}}=0$ if
vertices $v_{i}$ and $v_{j}$ are not neighbors, and ${{(A)}_{ii}}=0$
(undirected, unweighted, without loops and multiple edges graph) \cite{Mi 1}.

For the general case let us assume that $K$ is an oriented simplicial
complex, $q$ is an integer with $0<q\leq {dim(K)}$, and let $\{{\sigma ^{1}},%
{\sigma ^{2}},...,{\sigma ^{n}}\}$ denote the $q$-simplices of complex $K$,
then

\begin{equation*}
{({\mathcal{L}_{q}})_{ij}}=%
\begin{cases}
{deg_{U}}({\sigma ^{i}})+q+1, & \text{{\small {if $i=j$}}} \\
1, & \text{{\small {if $i\neq j$ and ${\sigma ^{i}}$ and ${\sigma ^{j}}$ are}%
}} \\
& \text{{\small {not upper adjacent but}}} \\
& \text{{\small {have a similar common}}} \\
& \text{{\small {lower simplex}}} \\
-1, & \text{{\small {if $i\neq j$ and ${\sigma ^{i}}$ and ${\sigma ^{j}}$ are%
}}} \\
& \text{{\small {not upper adjacent but}}} \\
\label{c2:lq} & \text{{\small {have a dissimilar common}}} \\
& \text{{\small {lower simplex}}} \\
0\ , & \text{{\small {if $i\neq j$ and ${\sigma ^{i}}$ and ${\sigma ^{j}}$
are}}} \\
& \text{{\small {upper adjacent or are not}}} \\
& \text{{\small {lower adjacent}}}%
\end{cases}%
.
\end{equation*}%

\section{The Model}

Opinions are modeled as simplices and and social interaction mechanisms were
presented in the Introduction. For simplicity, we have chosen that initially
each opinion is characterized by the same number of judgements implying that
each simplex has the same dimension $q_{in}$ at the beginning of the
simulation, although the case when each simplex has different dimension can
be easily implemented. The dimension $q_{in}$ as well as the total number of
judgements (i.e. vertices) and number of opinions associated with agents
will be changed due to interaction mechanisms. The relationship between
simplices (opinions) and vertices (judgements) is captured in the so called
incidence matrix $\Lambda $ \cite{Maletic}, \cite{Johnson}, in which rows
are associated to simplices and columns to vertices, such that the matrix
element is ${\Lambda _{ij}}=1$ if vertex $j$ is part of the simplex $i$, and
$0$ otherwise. From this matrix all other quantities describing the
structure of the interaction may be derived. During the simulation the size
of incidence matrix changes since vertices (judgements) can appear or
disappear and simplices (opinions) may change and also emerge or dissolve.
The actual stages of the algorithm are presented in the chronological order:

(1) The agents are located on the sites of the scale-free complex network
(graph).

(2) Fix two parameters, initial dimension of simplices $q_{in}$ and initial
number of judgements $v_{in}$ (${q_{in}}\leq {v_{in}}$).

(3) To each agent associate ${q_{in}}+1$ different random numbers between $1$
and $v_{in}$. Each different unordered set of ${q_{in}}+1$ judgements
defines a single simplex, and associated to each agent is an opinion defined
by the same number of judgements.

(4) Form initial simplicial complex of opinions $O_{in}$ (the initial
incidence matrix) from different unordered sets generated at step (3).
Calculate initial $Q$-vector, $\mathbf{Q}_{in}$. If all entries of $\mathbf{Q%
}_{in}$ are equal to $1$ simulation stops, otherwise continue to step (5).

(5) Randomly choose one of the agents $i$. If the simplex of agent's opinion
${\sigma _{q}}({o_{i}^{1}})=\langle {v_{0}},{v_{1}},{v_{2}},...,{v_{q}}%
\rangle $ has the associated vertex significance equal to $1$, i.e. $vs({%
\sigma _{q}}({o_{i}^{1}}))=1$, then extend his (her) opinion by an
additional judgement not previously present in the initial set of judgements
so that a new opinion ${\sigma _{q+1}}({o_{i}^{2}})=\langle {v_{0}},{v_{1}},{%
v_{2}},...,{v_{q}},{v_{q+1}}\rangle $ is formed. As a result of this process
sets of opinions and judgements are enlarged by adding new simplex and new
judgement and the dimension of simplicial complex of opinions is increased
by $1$. If $vs({\sigma _{q}}({o_{i}^{1}}))<1$ nothing happens and the
algorithm advances to step (6).

(6) Randomly choose one of the agents $i$, and one of its nearest neighbors $%
j$. The opinion simplices of these two agents are ${\sigma _{q_{1}}}({%
o_{i}^{1}})$ and ${\sigma _{q_{2}}}({o_{i}^{2}})$. We distinguish between
two cases:

(a) If opinions (i.e. simplices) of agents $i$ and $j$ are the same (${%
\sigma _{q_{1}}}({o_{i}^{1}})={\sigma _{q_{2}}}({o_{i}^{2}})$) the agents
convince all their common nearest neighbors to adopt this opinion. The
result of this step is the change of opinions of the neighboring agents $i$
and $j$ without any alteration of the simplicial complex of opinions.

(b) If opinions (i.e. simplices) of agents $i$ and $j$ are not the same but
they overlap the result of their communication can be either compromise or
joint opinion. If the overlap assumes sharing ${f_{ij}}+1$ judgements, i.e.
simplices corresponding to opinions share $f_{ij}$-face, determine \textit{%
the degree} \textit{of} \textit{compromise }for each of the opinions. First
define the \textit{overlap degrees} from the aspect of opinions of agents $i$
and $j$ as ${\omega _{i}}={f_{ij}}/{q_{i}}$ and ${\omega _{j}}={f_{ij}}/{%
q_{j}}$ respectively, where $q_{i}$ and $q_{j}$ are dimensions of opinion
simplices of agents $i$ and $j$, respectively. Then the degree of compromise
is defined as:
\begin{equation}
\phi ={\frac{({\omega _{i}}+{\omega _{j}})}{2}}\,.
\end{equation}%
Generate random number $r$ between $0$ and $1$ from the uniform
distribution. If $r\leq \phi $ then the agents merge their opinions into a
new opinion of dimension equal to ${q_{i}}+{q_{j}}-{f_{ij}}$ ($>{q_{i}},{%
q_{j}}$) formed from the union of individual judgements. Associate this new
opinion with each of the agents (i.e. the agents embrace this new opinion)
and add this new simplex to the simplicial complex of opinions . If $r>\phi $
then the opinion of one of the agents is changed by removing a judgement
(not from the overlap set) and then adopting a randomly chosen judgement
(not from the overlap set) from the opinion of other agent. The results of
this process is the change of the opinion of one agent and the addition of
the new simplex into simplicial complex of opinions. The stochastic
criterion used at this stage favors the joint opinion.

At each simulation stage the $Q$-vector is evaluated and simulation stops
when all entries of $Q$-vector are equal to $1$. If this condition is
fulfilled the consensus will be achieved after long enough time. The value
equal to $1$ for all $Q$-vector elements implies that there is only one
connectivity class at each connectivity level. When new opinions are formed
they will appear in some of the already present connectivity class and the
number of opinions in connectivity classes decrease.

Since the opinion space is geometrically represented as simplicial complex
we define the pure geometrical states of the opinion space in the analogy
with the quantum mechanical pure states. Taking a $q$-Laplacian matrix as a
density matrix at dimension $q$, a simplicial complex of opinions is formed
by pure states if the following relation is satisfied:
\begin{equation*}
Tr({L_{q}^{n}})=Tr({[{L_{q}^{n}}]^{2}})\,,
\end{equation*}%
where $Tr(\quad )$ is trace of the matrix, and ${L_{q}^{n}}$ $(=A\cdot {L_{q}%
})$ is the properly normalized Laplacian matrix for dimension $q$. Choosing
the normalization constant of the Laplacian matrix as
\begin{equation*}
A={\frac{{{\sum_{i}}{n_{i}}{d_{i}}}}{{{\sum_{i}}{n_{i}}{d_{i}^{2}}}}}\,,
\end{equation*}%
where $d_{i}$ are diagonal elements and $n_{i}$ is multiplicity of the $%
d_{i} $-th diagonal element, the "trace" condition is satisfied. In this
case we say that simplicial complex is in the pure state if it is formed by
the collection of disconnected simplices. At the end of simulation we check
whether the opinion space consists of pure states or not.

\section{Results of simulation}

In order to examine the consequences of the interaction mechanisms on
changes of the opinion space, we have performed a simulation with $S$\
agents located at the sites of the scale-free network with well defined
communities, called modules \cite{Boccaletti}. In the construction of
clustered modular scale-free network we used the algorithm introduced in
\cite{Mitrovic} with the following model parameters: $M$ - mean number of
links per node, $P_{0}$ - probability of module formation, $\alpha $ -
rewiring parameter which provides higher (lower) clustering coefficient and $%
N$ the number of nodes. Numerical values of parameters are $M=5$, ${P_{0}}%
=0.007$, $\alpha =0.6$, and $N=1000$. This network has high clustering
coefficient and $7$ modules. In the opinion space initial dimension of each
simplex (opinion) is ${q_{in}}=4$ and we varied the initial number of
vertices (judgements) from the set ${v_{in}}=[4,5,6,7,8]$. The focus of our
interest lies in monitoring changes in the opinion space regarding the
number of opinions associated to agents $S$ (some opinions disappear and
some new opinions are added), the number of judgements $v$ (new judgements
are added), the opinion $o_{m}$\ associated to the maximal number of agents $%
n_{m}$ , and topological and geometric properties of simplex associated with
the opinion $o_{m}$ such as eccentricity and vertex significance of simplex.
The maximal number of different realized opinions associated to agents
during the simulation is denoted by $S_{max}$, and the number of different
opinions associated to agents at the end of simulation is denoted by $%
S_{end} $. We recall that the simulation stops when consensus state is
reached which in the topology of the opinion space implies that the
simplicial complex of opinions is geometrically connected at all levels of
connectivity (recall the structure of the $Q$-vector) . We have tested this
assumption for all values from the set ${v_{in}}$ and in each case the
consensus was reached.

Since the initial number of judgements ${v_{in}}$ was changed in simulations
the dependence of $S_{max}$ and $S_{end}$ on the ratio ${q_{in}}/{v_{in}}$
was tested and is presented at Fig. 5 left and Fig. 5 right, respectively.
Both ${S_{max}}({\frac{{q_{in}}}{{v_{in}}}})$ and ${S_{end}}({\frac{{q_{in}}%
}{{v_{in}}}})$ relationships are in good agreement with the power law fit
with exponents $6.24$ and $5.86$, respectively. Their ratio $5.86/6.24=0.93$
is equal to the exponent of the power law dependence ${S_{end}}({S_{max}})$
which means that very small portion of opinions disappeared, primarily by
the convincing process. After reaching maximum the convincing process
assumes more important role so that some opinions disappear and connectivity
of opinion space at all levels of connectivity decreases.

\begin{center}
\includegraphics[width=80mm]{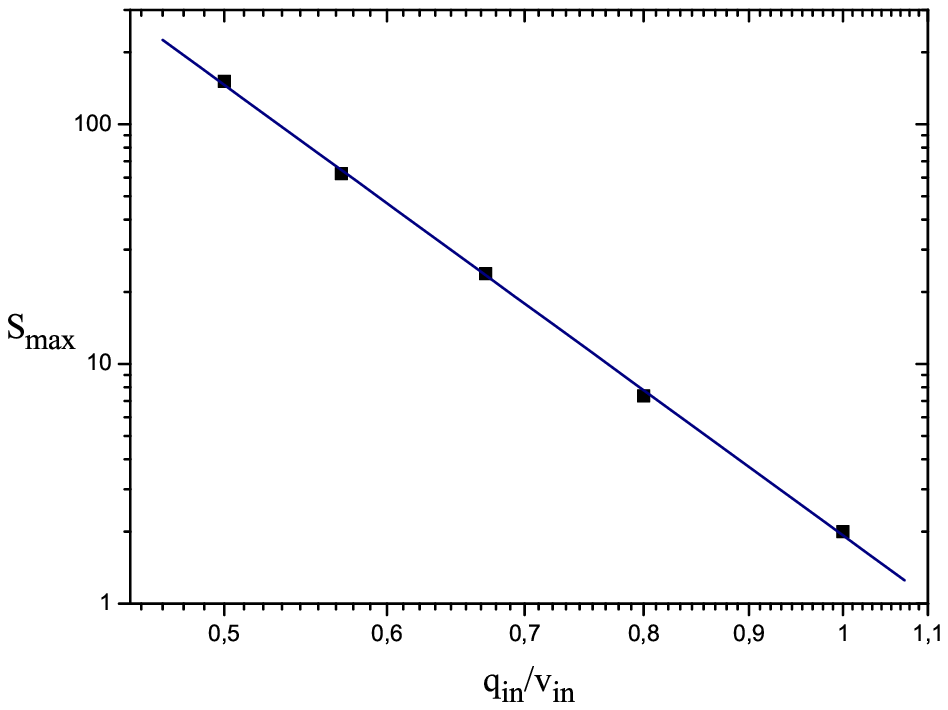}\hskip1mm %
\includegraphics[width=80mm]{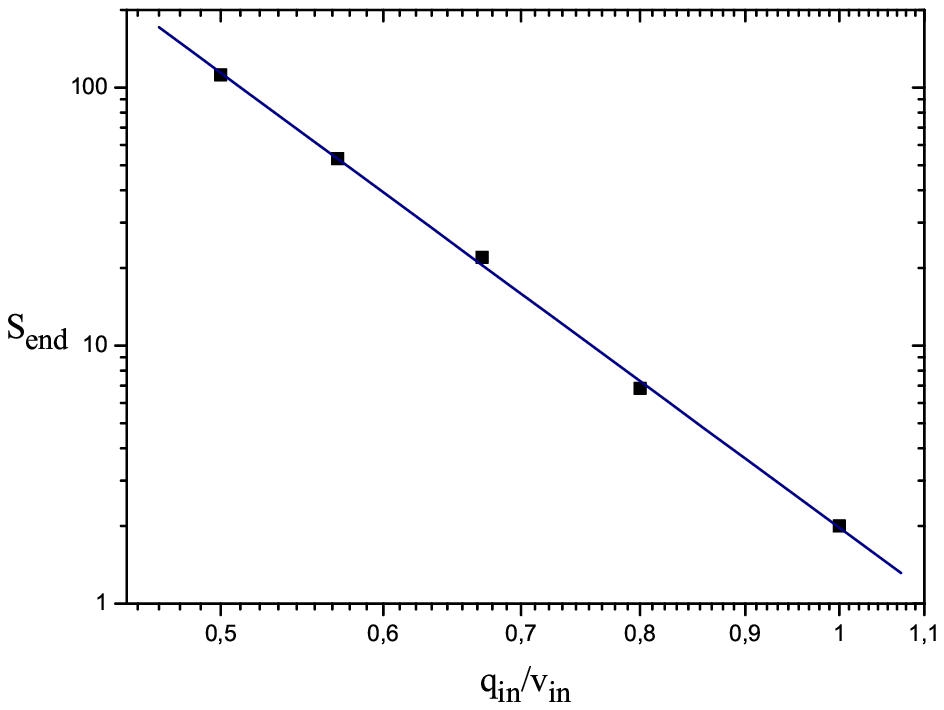}

Figure 5 Maximal number of different realized opinions associated to agents
during the simulation ($S_{max}$) (left), and the number of different
opinions associated to agents at the end of simulation ($S_{end}$) (right) as a function of the parameter ratio ${q_{in}}/{v_{in}}$.
\end{center}

The relationship between initial number of vertices (judgements) $v_{in}$
and the number of vertices (judgements) at the end of the simulation $%
v_{end} $ obeys a power law with exponent $1.1$ shown in Fig. 6. This value
of exponent suggests small influence of newly formed judgements on topology
of the opinion space.

\begin{center}
\includegraphics[width=80mm]{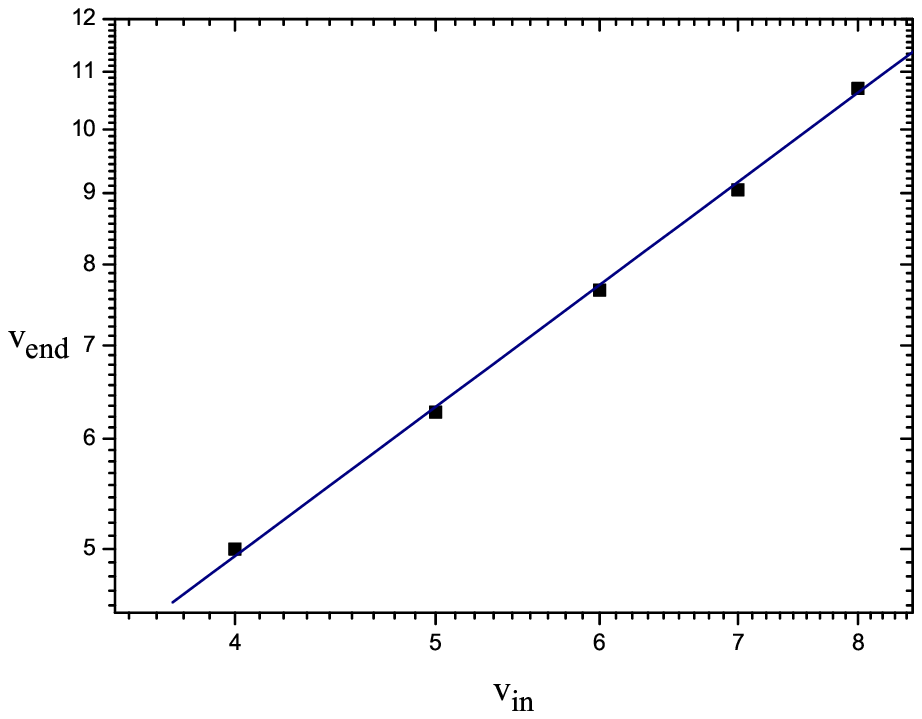}

Figure 6 Relationship between initial number of vertices (judgements) $v_{in}$
and the number of vertices (judgements) at the end of the simulation $v_{end}
$.
\end{center}

Fig. 7 displays dependence of the maximal number of agents $n_{m}$
associated with the single opinion on the initial parameter ${q_{in}}/{v_{in}%
}$. The relationship displays the power law with the exponent $-0.82$. It
turns out that keeping $q_{in}$ constant and increasing $v_{in}$ and hence
the number of different opinions, the largest number of agents associated to
the single opinion$\ n_{m}$ grows. This is somewhat paradoxical since it
implies that in spite of the growing number of opinions and hence choices
for the agents the number of individuals adhering to and/or adopting the
same opinion increases.

\begin{center}
\includegraphics[width=80mm]{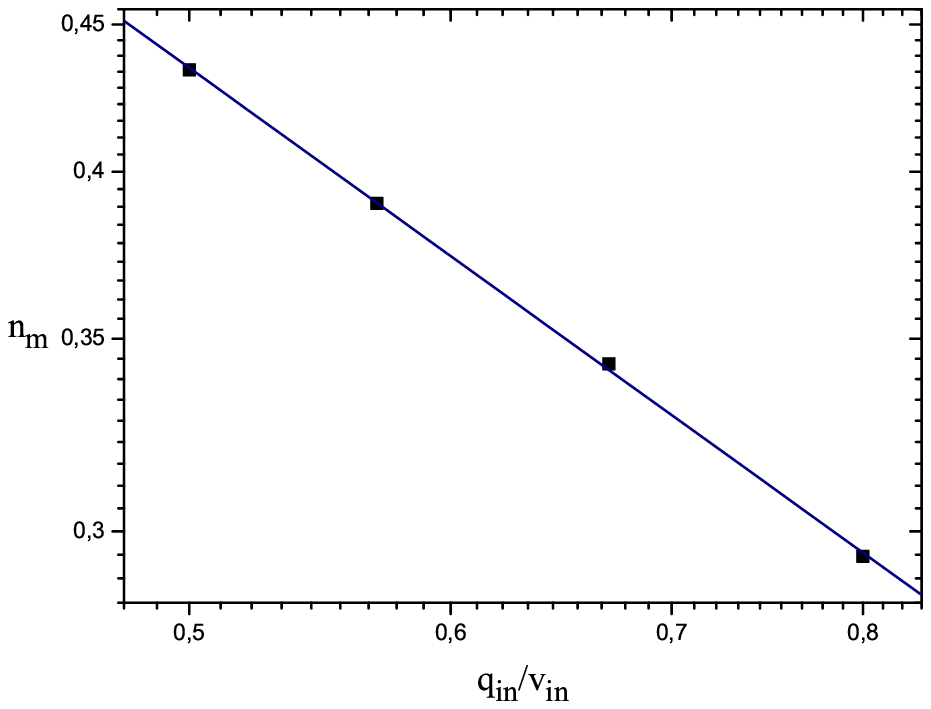}

Figure 7 Dependence of the maximal number of agents $n_{m}$
associated with the single opinion on the initial parameter ratio ${q_{in}}/{v_{in}}$.
\end{center}

At the end of the simulation of special interest is the eccentricity and
vertex significance of the opinion $o_{m}$ adopted by the largest number of
agents. The eccentricity value for all $v_{in}$ is around $0$, which means
that this opinion is well integrated into the opinion structure, and that it
shares almost all judgements with the other opinions. The vertex
significance value for all $v_{in}$ is in the range $8.1\leq vs\leq 9.1$
confirming the importance of the opinion $o_{m}$\ with the largest number of
individuals who embrace it. The fact that it shares all of its judgements
with other opinions but still attracts the largest number of individuals
bolsters its importance. The opinion space in all considered cases consists
of pure states at the end of the simulation.

\section{Conclusion}

An opinion dynamics model which assumes opinions as unordered sets of
judgements is developed. The opinion space is mapped to simplicial complex
so that the analysis uses concepts of combinatorial algebraic topology,
primarily relying on the Q-vector. Through the interaction and communication
mechanisms opinions appear and disappear causing changes in the topology of
opinion space monitored by evolution of the Q-vector. When all entries of
Q-vector are equal to $1$ the consensus is achieved.

We have calculated the maximal number of realized opinions associated to
agents during the simulation, the number of realized different opinions
associated to agents at the end of simulation (when entries of Q-vector are
equal to $1$), number of judgements at the end of the simulation, and the
largest number of agents associated to the single opinion. All these
quantities display power law dependence as functions of $v_{in}$ where $%
v_{in}$ is the initial number of vertices. Particulary interesting is the
increase of the largest number of agents associated to the same single
opinion as a function of increasing $v_{in}$. It turns out that the larger
the initial difference between opinions the larger number of agents are
associated with the same opinion, which is in contradiction with the results
for the Axelrod model \cite{Axelrod}. These results are valid for only one
fixed value of $q_{in}$ and five small values of $v_{in}$ and the future
work will be include the change of these two initial values and examination
whether the above quantities follow the same behavior. The influence of
external factors such as the mass-media or marketing may be easily
introduced in the model by representing them as an external simplicial
complex. The opinion associated with the largest number of agents adhering
to the same opinion is well integrated into the structure of opinions and it
would be interesting to check whether the consensus is achieved over this
opinion. The results presented here indicate that at the end of simulation
the opinion space is formed by the pure states and that the future opinion
dynamics is strongly dominated by the convincing process. The intention of
the authors was not to predict any real situation with this model but to
highlight coupling of different communication mechanisms and the structure
and evolution of the opinion space through the applications of the
simplicial complex.

\end{document}